\documentclass[twocolumn,aps,showpacs,nofootinbib,epsfig]{revtex4}
\usepackage{amssymb}
\usepackage{amsmath}
\usepackage{graphics}
\usepackage{epsfig}
\usepackage[dvips]{color}

\newcommand{\be}{\begin{equation}}
\newcommand{\ee}{\end{equation}}
\newcommand{\bea}{\begin{eqnarray}}
\newcommand{\eea}{\end{eqnarray}}
\newcommand{\beas}{\begin{eqnarray*}}
\newcommand{\eeas}{\end{eqnarray*}}

\newcommand{\slsh}[1]{{\not \! #1}}

\begin{document}
\title{QCD phase diagram from finite energy sum rules}
\author{Alejandro Ayala$^1$, Adnan Bashir$^2$, C. A. Dominguez$^3$, Enif
  Guti\'errez$^2$, M. Loewe$^4$ and Alfredo Raya$^2$}
\affiliation{$^1$Instituto de Ciencias Nucleares, Universidad Nacional
Aut\'onoma de M\'exico, Apartado Postal 70-543, M\'exico Distrito Federal
04510, Mexico.\\ 
$^2$Instituto de F\1sica y Matem\'aticas,
Universidad Michoacana de San Nicol\'as de Hidalgo, Edificio C-3, Ciudad
Universitaria, Morelia, Michoac\'an 58040, Mexico.\\ 
$^3$Centre for Theoretical Physics and Astrophysics, University of Cape Town,
Rondebosch 7700, South Africa and Department of Physics, Stellenbosch
University, Stellenbosch 7600, South Africa.\\ 
$^4$Pontificia Universidad Cat\'olica de Chile, Facultad de F\1sica, Casilla
306, Santiago 22, Chile.}

\begin{abstract}

We study the QCD phase diagram at finite temperature and
baryon chemical potential by relating the behavior of the
light-quark condensate to the  threshold energy for  the onset of perturbative
QCD. These parameters are connected to the chiral symmetry restoration and the
deconfinement phase transition, respectively. This relation is obtained in the
framework of finite energy QCD sum rules at finite temperature and density,
with input from Schwinger-Dyson methods to determine the light-quark
condensate. Results indicate that both critical temperatures are basically the
same within some $3 \%$ accuracy. We also obtain bounds for the position of
the critical end point, $\mu_{B c} \gtrsim 300$ MeV and $T_c\lesssim 185$
MeV.

\end{abstract}


\maketitle

\section{Introduction}\label{I}
In Quantum Chromodynamics (QCD) the strong interaction among quarks depends on
their color charge. When quarks are placed in a medium this color charge is
screened with increasing density. The density can increase either by raising
the temperature, so that collisions between quarks produce more quarks and
gluons, or by compressing the system, thereby increasing the baryon
density. If the density increases beyond a certain critical value one expects
that the interactions between quarks no longer confine them inside a hadron,
so that they are free to travel longer distances and deconfine.  This
transition from a confined to a deconfined phase is usually  referred to as
the {\it deconfinement phase transition.} 

A separate phase transition takes
place when the realization of chiral symmetry shifts from a Nambu-Goldstone
phase to a Wigner-Weyl phase. In the massless quark limit this is achieved by
the vanishing of the quark condensate, or alternatively the pion decay
constant. Qualitatively, one expects these two phase transitions to take place
at approximately the same temperature. An outstanding 
issue is whether this conclusion also holds quantitatively. To address this,
it has been customary to study the behavior of the
order parameters of these transitions as functions of the temperature $T$ and
the baryon chemical potential $\mu_B$, namely the Polyakov loop
$L$~\cite{Larry2} and quark anti-quark condensate $\langle\bar{\psi}
\psi\rangle$ in the chiral limit, respectively. In the
confined phase the former parameter either vanishes in the limit of massless
quarks, or else it is exponentially suppressed 
for finite quark masses, while it is finite in the deconfined phase. The
quark condensate is finite in the confined phase, while it vanishes in the deconfined phase at high enough temperature, and
in the limit of massless quarks. 
For finite quark masses chiral symmetry is explicitly broken at the Lagrangian
level and therefore the phase transition is suppressed. This is similar to
what happens to a ferromagnet in the presence of an external magnetic
field. In this situation one might need to specify to what extent one is still
dealing with a phase transition.

At finite $T$, and $\mu_B =0$, lattice QCD calculations provide a
consistent quantitative picture of the above behavior, resulting in similar critical temperatures $T_c$ for both transitions in the range $170$ MeV
$\lesssim T_c \lesssim$ $200$ MeV, for finite quark masses~\cite{latticeT,
latticelatest, latticerev}.
The situation is much less clear cut when both $T$ and $\mu_B$ are simultaneously non-zero.
Lattice QCD simulations cannot be used for $\mu_B\neq 0$ because the fermion
determinant becomes complex and thus standard Monte Carlo methods fail, as the
integrand is no longer real and positive definite. However, these techniques
can still be adapted to extract some, though not exact, information on the QCD
phase diagram for $\mu_B\neq 0$~\cite{Karsch}. Therefore, one  needs to resort either to mathematical constructions to overcome the above
limitation~\cite{Alexandru-Endrodi}, or to model calculations~\cite{Sousa}. Of
particular recent interest is the search for a possible critical
end point~\cite{CEP} that signals the strengthening of the order of the
transition with increasing $\mu_B$, indicating the beginning of a true chiral
symmetry restoring/deconfining phase transition. 
The results from Monte Carlo simulations and model calculations, with and
without Polyakov loop, or its variants, seem to be in conflict. In fact, the
former give smaller (larger) values for the end point baryon chemical
potential (temperature) than the latter. 
Things become worse if one uses the imaginary chemical potential method,
a well established technique for not too large values of $\mu_B$. Indeed, this leads to  a shrinking and weakening region of chiral phase transitions with increasing $\mu_B$, thus suggesting that there is no critical end point for $\mu_B\lesssim 500$ MeV ~\cite{Philipsen}. It has
also been pointed out that even if the transition weakens with increasing
$\mu_B$, the existence of the critical end point would not be ruled out,
although it would require a non-monotonic behavior ~\cite{Kapusta}. In view of
this situation, alternative ways of examining the QCD phase diagram are
required. 

One possibility is to look at variables that describe
deconfinement other than the Polyakov loop. A phenomenological QCD parameter associated with deconfinement was first proposed long ago in \cite{pioneer}, and it is the square energy ($s_0$) beyond which the hadronic resonance spectral function becomes smooth and well described by perturbative QCD (PQCD). At $T=0$ this continuum threshold lies in the range $s_0 \simeq 1 - 3 \; \mbox{GeV}^2$, depending on the channel.
At finite temperature one expects $s_0$ to decrease with increasing $T$ and
approach the kinematical threshold at some critical value $T = T_c$, to be
identified with the deconfinement temperature. In this scenario one expects
stable particles (poles on the real axis in the complex squared energy
$s$-plane) to develop a width as a result of absorption in the thermal
bath. At the same time, resonances (poles in the second Riemann sheet in the
complex $s$-plane) should develop $T$-dependent widths, increasing with
increasing temperature. Such a  {\it resonance broadening} mechanism was first
proposed in detail in connection with dimuon production in heavy ion
collisions \cite{dimuon}. 

The natural framework to determine $s_0$ has been that of QCD sum rules \cite{QCDSRreview}. This quantum field theory framework is based on the operator product expansion (OPE) of current correlators at short distances, extended beyond perturbation theory, and on Cauchy's theorem in the complex $s$-plane. The latter is usually referred to as quark-hadron duality. Vacuum expectation values of quark and gluon field operators effectively parametrize the effects of confinement. An extension
of this method to finite temperature was first outlined in \cite{pioneer}. Further evidence supporting the validity of this program was provided in \cite{OPET}, followed by a large number of applications \cite{Dominguez}-\cite{QCDT}. Of particular interest to the  present work are the results obtained  for $s_0(T)$ in \cite{Dominguez} using QCD Finite Energy Sum Rules (FESR) for the (light-quark) axial-vector current correlator. The leading dimension FESR relates $s_0(T)$ to the pion decay constant $f_\pi(T)$, and this in turn to the light-quark condensate (using the Gell-Mann-Oakes-Renner relation \cite{Dominguez2}). In the chiral limit it was found that $s_0(T)/s_0(0) \simeq f_\pi(T)/f_\pi(0) \simeq \langle\bar{\psi} \,\psi\rangle(T)/\langle\bar{\psi} \,\psi\rangle(0)$, which holds to a very good approximation. This relation hints towards the possible coincidence of the critical temperatures for deconfinement and chiral-symmetry restoration. In this paper we extend this analysis to finite density, thus obtaining $s_0(T,\mu_B)$ from FESR using as input the light-quark condensate at finite temperature and density determined in the Schwinger-Dyson equations (SDE) framework.\\

The paper is organized as follows: In Sec.~\ref{II} we find the relation between the quark condensate and the PQCD threshold $s_0$ using FESR for the axial-vector current correlator. In Sec.~\ref{III} we compute the quark condensate at finite $T$ and
$\mu_B$ from a convenient parametrization of the quark propagator in the SDE framework. In Sec.~\ref{IV} we present our analysis of the QCD phase diagram and show that
 the deconfinement and chiral symmetry restoration transitions take place at basically the same temperature to some $3\,\%$ accuracy, i.e.
within the numerical precision of the method. We finally
summarize and discuss our results in Sec.~\ref{V}.  

\section{Finite energy QCD sum rules}\label{II}

We begin by considering  the (charged) axial-vector current correlator at $T=0$
\begin{eqnarray}
   \Pi_{\mu\nu}(q^2) &=& i\int d^4x \,e^{iq\cdot x}\,
   \langle 0| T(A_\mu(x) A_\nu(0))|0 \rangle, \nonumber \\ [.3cm]
   &=& - g_{\mu\nu} \, \Pi_1(q^2) + q_\mu q_\nu \Pi_0(q^2) \;,
\label{correlator}
\end{eqnarray}
where $A_\mu(x) = :\bar{u}(x) \gamma_\mu \gamma_5 d(x):$ is the axial-vector current, $q_\mu = (\omega, \vec{q})$ is the four-momentum transfer, and the functions $\Pi_{0,1}(q^2)$ are free of kinematical singularities. Concentrating on the function $\Pi_0(q^2)$ and writing the OPE beyond perturbation theory in QCD \cite{QCDSRreview}, one of the two pillars of the sum rule method,  one has
\begin{equation}
\Pi_0(q^2)|_{\mbox{\tiny{QCD}}} = C_0 \, \hat{I} + \sum_{N=1} C_{2N} (q^2,\mu^2) \langle \hat{\mathcal{O}}_{2N} (\mu^2) \rangle \;, \label{OPE}
\end{equation}
where $\mu^2$ is a renormalization scale, the Wilson coefficients $C_N$ depend on the Lorentz indices and quantum numbers of the currents and on the local gauge invariant operators ${\hat{\mathcal{O}}}_N$ built from the quark and gluon fields in the QCD Lagrangian. These operators are ordered by increasing dimensionality and the Wilson coefficients, calculable in PQCD, fall off by corresponding powers of $-q^2$. The unit operator above has dimension $d=0$ and $C_0 \hat{I}$ stands for the purely perturbative contribution. Hence, this OPE factorizes short distance physics, encapsulated in the Wilson coefficients, and long distance effects parametrized by the vacuum condensates. The second pillar of the QCD sum rule technique is Cauchy's theorem in the complex squared energy $s$-plane
\begin{equation}
\frac{1}{\pi}\int_{0}^{s_0} ds f(s) \mbox{Im} \Pi_0 (s)
   =
   -\frac{1}{2\pi i}\oint_{C(|s_0|)}ds f(s) \Pi_0 (s) \;,
   \label{disprel}
\end{equation}
where $f(s)$ is an arbitrary analytic function, and the radius of the circle $s_0$ is large enough for QCD and the OPE to be used on the circle (see Fig.1). The integral along the real $s$-axis involves the hadronic spectral function. This equation is the mathematical statement of what is usually referred to as {\it quark-hadron duality}. Using the OPE, Eq.(\ref{OPE}), and an integration kernel $f(s) = s^N \; (N=1,2,\cdots)$ one obtains the FESR
\begin{eqnarray}
(-)^{N-1} &C_{2N}& \langle {\mathcal{\hat{O}}}_{2N}\rangle = 4 \pi^2 \int_0^{s_0} ds\, s^{N-1} \,\frac{1}{\pi} {\mbox{Im}} \Pi_0(s)
\nonumber \\ [.3cm]
&-& \frac{s_0^N}{N} \left[1+{\mathcal{O}}(\alpha_s)\right] \;\; (N=1,2,\cdots) \;.\label{FESR}
\end{eqnarray}
\begin{figure}[t!] 
\resizebox{0.4\textwidth}{!}{ \includegraphics{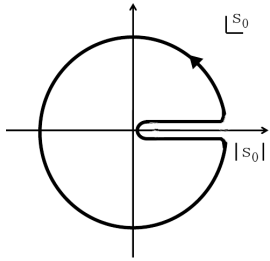}}
\caption{FESR integration contour $C(|s_0|)$ in the complex square energy
  $s$-plane. The QCD threshold $s_0$ in the FESR is the  radius of the
  circle.}
\label{fig1}
\end{figure}
For $N=1$ the dimension $d=2$ term in the OPE does not involve any condensate, as it is not possible to construct a gauge invariant operator of such a dimension from the quark and gluon fields. Nevertheless, it is a-priori conceivable to generate a $d=2$ term in some dynamical fashion, e.g. in  PQCD at very high order (renormalons). However, there is no evidence for
such a term (at $T=0$) from FESR analyses of experimental data on $e^+ e^-$ annihilation and $\tau$ decays into hadrons \cite{C2}. At very high temperatures, though, there seems to be evidence for some $d=2$ term \cite{C2T}. However, the analysis to be reported here is performed at much lower values of $T$, so that we can safely ignore this contribution in the sequel.\\

The extension of this program to finite temperature is fairly straightforward \cite{pioneer}, \cite{OPET}, with the Wilson coefficients in the OPE, Eq.(\ref{OPE}), remaining independent of $T$ at leading order in $\alpha_s$, and the condensates developing a temperature dependence. Radiative corrections in QCD  involve now an additional scale, i.e. the temperature, so that $\alpha_s \equiv \alpha_s(\mu^2,T)$. This problem has not yet been  solved successfully. Nevertheless, from the size of radiative corrections at $T=0$ one does not expect any major loss of accuracy in results from thermal FESR to leading order in PQCD, as long as the temperature is not too high, say $T \lesssim 200 \, {\mbox {MeV}}$. Essentially all applications of  FESR at $T \neq 0$ have been done at leading order in PQCD, thus implying a systematic uncertainty at the level of 10 \%. One new feature at $T \neq 0$ is the appearance of a new cut in the complex energy $\omega$-plane \cite{pioneer}, and centered at the origin with extension $- |\vec{q}| \leq \omega \leq |\vec{q}|$. This is due to a contribution to the current correlator in the space-like region ($q^2 < 0$) which vanishes at $T=0$. Conceptually, this originates  in the scattering of the current  by either quarks (antiquarks) or by hadrons in the medium, in the case of QCD or the hadronic representation, respectively. When considering the rest-frame ($\vec{q} \rightarrow 0$)
this scattering term either becomes a delta function of the energy or it vanishes identically, depending on the channel. For instance, in the case of the axial-vector current correlator, Eq.(\ref{correlator}), the QCD scattering term is proportional to $\delta(\omega^2)$. The corresponding term in the hadronic representation is non-zero but it is  suppressed relative to the tree-level pion contribution, as the axial-vector current can only couple to an odd number of pions. Another new feature at finite temperature is the possible existence of non-scalar (Lorentz noninvariant) vacuum condensates. This does not affect the present analysis, as we shall only consider dimension $d=2$ FESR.\\

In the static limit ($\vec{q} \rightarrow 0$), to leading order in PQCD, and for $T\neq 0$ and $\mu_B \neq 0$  the  function $\Pi_0(q^2)$ in Eq.(\ref{correlator}) becomes
$\Pi_0(\omega^2, T, \mu_B)$; to simplify the notation we shall omit the $T$ and $\mu_B$ dependence in the sequel. A straightforward calculation of the spectral function in perturbative QCD gives
\begin{eqnarray}
   &\frac{1}{\pi}&{\mbox{Im}}\Pi_0(s)|_{\mbox{\tiny{PQCD}}}
   =
   \frac{1}{4\pi^2}\left[1-\tilde{n}_+\left(\frac{\sqrt{s}}{2}\right) 
   -\tilde{n}_-\left(\frac{\sqrt{s}}{2}\right)\right] \nonumber \\[.3cm]
   &-&\frac{2}{\pi^2} \;T^2 \;\delta (s)\; \left[
   {\mbox{Li}}_2(-e^{\mu_B/T})
   + {\mbox{Li}}_2(-e^{-\mu_B/T})\right],
\label{pertQCD}
\end{eqnarray}
where ${\mbox{Li}}_2(x)$ is the dilogarithm function, $s=\omega^2$, and
\begin{equation}
   \tilde{n}_\pm(x)=\frac{1}{e^{(x\mp \mu_B)/T}+1}
\label{F-D}
\end{equation}
are the Fermi-Dirac thermal distributions for particles and antiparticles,
respectively. We have assumed massless quarks, as quark mass corrections are negligible. However, later when we determine the quark condensate in Section III this approximation will be relaxed.

In the limit where $T$ and/or $\mu_B$ are large, Eq.~(\ref{pertQCD}) becomes
\begin{eqnarray}
   \frac{1}{\pi}{\mbox{Im}}\Pi_0(s)|_{\mbox{\tiny{PQCD}}}
   &=&
   \frac{1}{4 \pi^2}\left[1-\tilde{n}_+\left(\frac{\sqrt{s}}{2}\right) 
   -\tilde{n}_-\left(\frac{\sqrt{s}}{2}\right)\right]\nonumber\\ [.3cm]
   &+&\frac{1}{\pi^2}\;\delta (s)
   \left(\mu_B^2 + \frac{\pi^2T^2}{3} \right)\;.
\label{pertQCDlargeTmu}
\end{eqnarray}
\begin{figure}[t!] 
\resizebox{0.5\textwidth}{!}{ \includegraphics[angle=-90]{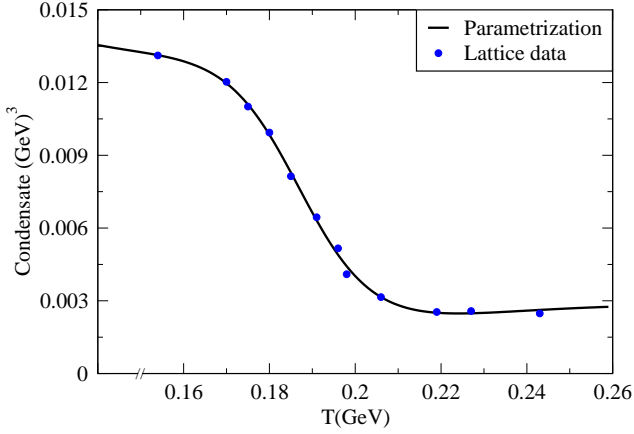}}
\caption{(Color online) Lattice data and parametrization of the absolute value
  of the quark condensate as a function of $T$ in the phase transition region.}
\label{fig2}
\end{figure}
In the hadronic sector we assume pion-pole dominance of the hadronic spectral function, i.e. the continuum threshold $s_0$ to lie below the first radial excitation with mass $M_{\pi_1} \simeq 1 300\;{\mbox{MeV}}$. This is a very good approximation at finite $T$, as we expect $s_0$ to be monotonically decreasing with increasing temperature. In this case, 
\begin{equation}
   \frac{1}{\pi}{\mbox{Im}}\Pi (s)
   |_{\mbox{\tiny{HAD}}}
   = 2 \; f_\pi^2(T,\mu_B)\; \delta (s),
\label{HAD}
\end{equation}
where $f_\pi(T,\mu_B)$ is the pion decay constant at finite $T$ and $\mu_B$,
with $f_\pi(0,0) =92.21 \pm 0.14 \;{\mbox{MeV}}$ \cite{PDG}.\\

Turning to the FESR, Eq.(\ref{FESR}), with $N=1$ and no dimension $d=2$ condensate, and using Eqs.(\ref{pertQCD}) and (\ref{HAD}) one finds
\begin{eqnarray}
&\frac{}{}&\int_0^{s_0(T,\mu_B)} ds \, \left[1 - \tilde{n}_+\left(\frac{\sqrt{s}}{2}\right) 
   -\tilde{n}_-\left(\frac{\sqrt{s}}{2}\right)\right] = \nonumber\\ [.3cm]
   &&8 f_\pi^2(T,\mu_B) - 8 T^2\left[{\mbox{Li}}_2(-e^{\mu_B/T})
   + {\mbox{Li}}_2(-e^{-\mu_B/T})\right]\frac{}{} \label{FESRTMU}
\end{eqnarray}
This is a transcendental equation determining $s_0(T,\mu_B)$ in terms of $f_\pi(T,\mu_B)$. The latter is related to the light-quark condensate through the Gell-Mann-Oakes-Renner relation \cite{Dominguez2}
\begin{equation}
\frac{f_\pi^2(T,\mu_B)}{f_\pi^2(0,0)} = \frac{\langle\bar{\psi}\psi\rangle(T,\mu_B)}{\langle\bar{\psi}\psi\rangle(0,0)}\;, \label{GMOR}
\end{equation}
where the quark and pion masses have been assumed independent of $T$ and $\mu_B$ \cite{Altherr}. A good closed form approximation to the FESR, Eq.(\ref{FESRTMU}), for large $T$ and/or $\mu_B$ is obtained using Eq.(\ref{pertQCDlargeTmu}) with $\tilde{n}_+\left(\frac{\sqrt{s}}{2}\right) \simeq \tilde{n}_-\left(\frac{\sqrt{s}}{2}\right) \simeq 0$, in which case
\begin{equation}
   s_0(T,\mu_B) \simeq  8 \, \pi^2\,
   f_\pi^2(T,\mu_B)
   -
   \frac{4}{3} \,\pi^2\, T^2 - 4 \, \mu_B^2\;.
\label{s0fpi}
\end{equation}
Using Eq.(\ref{GMOR}) this can be rewritten as
\begin{equation}
  \frac{s_0(T,\mu_B)}{s_0(0,0)} \simeq \frac{\langle\bar{\psi}\psi\rangle (T,\mu_B)}{\langle\bar{\psi}\psi\rangle (0,0)}
   -
   \frac{(T^2/3 - \mu_B^2/\pi^2)}{2 f_\pi^2(0,0)}
\label{s0cond}
\end{equation}
The quark condensate can be computed from the in-medium quark propagator,
whose non-perturbative properties can be obtained e.g. from known
solutions to the Schwinger-Dyson equations (SDE) as discussed in the next section.

\section{Quark propagator and condensate}\label{III}

The quark condensate can be computed from the quark propagator
$S(k_0,\vec{k})$ in Euclidean space. At finite $T$ and $\mu_B$ the condensate
is given by 
\begin{eqnarray}
  && \langle\bar{\psi}\psi\rangle (T,\mu_B)= -
   N_cT\sum_n\int \frac{d^3k}{(2\pi)^3}\nonumber\\[.3cm]
   &\times&{\mbox{Tr}}\ S[(2n+1)\pi T+i\mu_B , \vec{k}]\nonumber\\[.3cm]
   &=&-
   N_cT\sum_n\int \frac{d^4k}{(2\pi)^3}
   {\mbox{Tr}}[S(k_0,\vec{k})]\nonumber\\[.3cm]
   &\times&\delta [k_0 - (2n+1)\pi T -i\mu_B ].
\label{condensatefromprop}
\end{eqnarray}

\begin{table}[t!] 
\centering
\begin{tabular}{ccc}
\hline
   $i$ & $m_i$ (GeV) & $r_i$\\ [0.5ex]
\hline
    1 &  -0.490      & -0.112   \\
    2 &   0.495      &  0.352   \\
    3 &  -0.879      &  0.259   \\ [1ex]
\hline
\end{tabular}
\caption{Parameters $m_i$ and $r_i$, $i=1,2,3$ to describe the Lorentz
  covariant part of the quark propagator.} 
\end{table}

Introducing the Poisson summation formula
\begin{eqnarray}
   &&\sum_l(-1)^l\exp\{(ik_0 + \mu_B )l/T\} =\nonumber\\
   &&(2\pi )\ T\sum_n\delta [k_0 - (2n+1)\pi T -i\mu_B ],
\label{Poisson}
\end{eqnarray}
leads to 
\begin{eqnarray}
   &&T\sum_n\int \frac{d^3k}{(2\pi)^3}
   {\mbox{Tr}}S[(2n+1)\pi T+i\mu_B , \vec{k}]
   =\nonumber\\
   &&\sum_l(-1)^l\int\frac{d^4k}{(2\pi )^4}
   {\mbox{Tr}}[S(k_0,\vec{k})]
   \exp\{(ik_0 + \mu_B )l/T\}\nonumber\\\;.
\label{usingPoisson}
\end{eqnarray}
Using this result in Eq.(\ref{condensatefromprop}) gives
\begin{eqnarray}
 &&  \langle\bar{\psi}\psi\rangle (T,\mu_B)=-
   N_c\sum_l(-1)^l\int \frac{d^4k}{(2\pi)^4}\nonumber\\
   &\times&{\mbox{Tr}}[S(k_0 , \vec{k})]\exp\{(iq_0 + \mu_B )l/T\}.
   \nonumber\\
\label{condensatePoisson}
\end{eqnarray}
\begin{figure}[t!] 
\resizebox{0.5\textwidth}{!}{ \includegraphics[angle=-90]{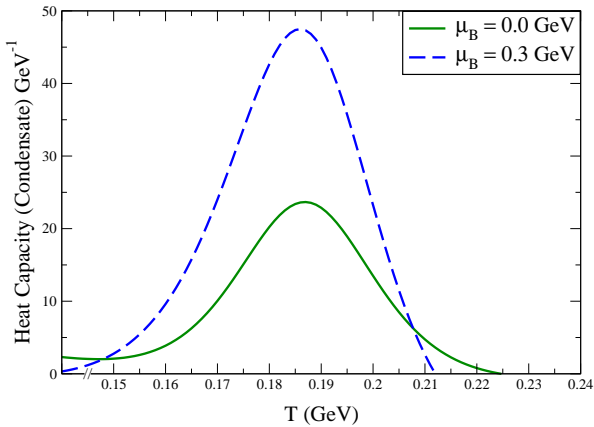}}
\caption{(Color online) Heat capacity for the quark condensate as a function
  of $T$ for $\mu_B =0$ (solid line) and $\mu_B= 300 \; {\mbox{MeV}}$ (dash
  line). The critical temperature $T_c$ corresponds to the maximum of the heat
  capacity for a given value of $\mu_B$. }  
 
\label{fig3}
\end{figure}
Notice that from Eq.~(\ref{condensatePoisson}) the vacuum contribution to the
condensate comes from the term with $l=0$. For this we use the value
\begin{equation}
   \langle\bar{\psi}\psi\rangle|_0=-(0.241\ {\mbox{GeV}})^3.
\label{condensatevacuum}
\end{equation}
The true matter contribution to the condensate is thus
\begin{eqnarray}
 &&  \langle\bar{\psi}\psi\rangle (T,\mu_B)=-
   N_c\sum_{l\neq 0}(-1)^l\int \frac{d^4k}{(2\pi)^4}\nonumber\\
   &\times&{\mbox{Tr}}[S(k_0 , \vec{k})]\exp\{(iq_0 + \mu_B )l/T\}.
   \nonumber\\
\label{truecondensate}
\end{eqnarray}
Due to the loss of Lorentz covariance at finite $T$ and/or $\mu_B$, the
general structure of the propagator is given by
\begin{equation}
   S^{-1}(k_0,\vec{k})=A \gamma_0 k_0 + B
   \vec{\gamma}\cdot\vec{k} + C,
\label{propgenfiniteTmu}
\end{equation}
where $A$, $B$ and $C$ are scalar functions of $k_0$ and $\vec{k}$. They can be
obtained from non-perturbative methods such as solutions to SDE. We adopt this procedure here. Motivated by the success of the
rainbow-ladder truncation of the SDE and the effective interaction of
Ref.~\cite{MT} in the description of light pseudo-scalar and vector mesons,
and the meromorphic representation of the quark propagator~\cite{Param}, we
consider the parametrization 
\begin{equation}
   S(k_0,\vec{k})=\sum_{i=1}^3\left(\frac{r_i}{i\slsh{k}+m_i}\right)+ 
   \frac{r_4}{i \gamma_0k_0 + i b\vec{\gamma}\cdot\vec{k}+m_4},
\label{parametrization}
\end{equation}
and choose $b$, the masses $m_i$, and the residues $r_i$,  $i=1\ldots 4$, to
be real numbers. In addition we seek $T$-dependent values of $b,\ m_4$ and
$r_4$. The Lorentz covariant part of this parametrization is fitted
by requiring the propagator to reproduce key features of the rainbow-ladder
model~\cite{MT} at $T=0$. In particular, to match the ultra-violet behavior of
the gap equation for massive $u/d$ quarks, the value of the condensate in
vacuum, and the constituent quark masses, as dictated by the of solutions to SDE. Table~I shows the values thus obtained for the
parameters $m_i$ and $r_i$,  $i=1\ldots 3$. The last term in
Eq.~(\ref{parametrization}) is added to reproduce the 
Lorentz covariance breaking effects of the heat bath at $T\neq 0$ and/or
$\mu_B\neq 0$. The values of $b$, $m_4$ and $r_4$ are adjusted to
reproduce the light-quark condensate as a function of $T$ for
$\mu_B=0$ [see Eq.~(\ref{condexpl}) below] extracted from 
lattice QCD~\cite{latticelatest} by means of a point-distance minimization
procedure.
\begin{figure}[t!] 
\resizebox{0.5\textwidth}{!}{ \includegraphics[angle=-90]{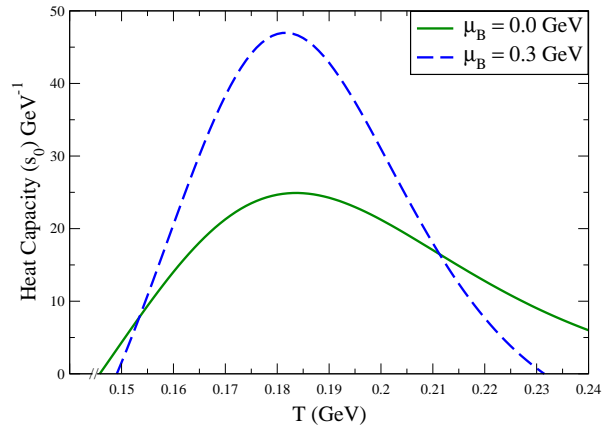}}
\caption{(Color online) Heat capacity for the PQCD threshold $s_0$ as a
  function of $T$ for $\mu_B =0$ (solid line) and $\mu_B= 300 \; {\mbox{MeV}}$
  (dash line). The critical temperature $T_c$ corresponds to the maximum of
  the heat capacity for a given value of $\mu_B$.}  
 
\label{fig4}
\end{figure}
\begin{figure}[b!] 
\resizebox{0.5\textwidth}{!}{ \includegraphics[angle=-90]{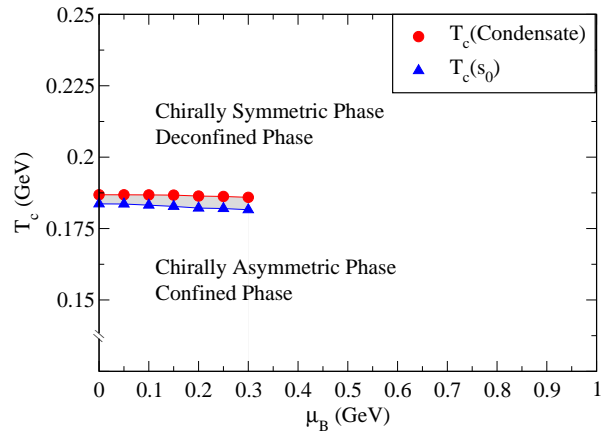}}
\caption{(Color online) Transition temperatures for the quark condensate and
  the PQCD
 threshold $s_0$  as functions of the baryon chemical potential.}
\label{fig5}
\end{figure}
Carrying out the integrations in Eq.~(\ref{truecondensate}), and in terms of
the parametrization of the quark propagator in Eq.~(\ref{parametrization}), we obtain
\begin{eqnarray}
   \langle\bar{\psi}\psi\rangle(T,\mu_B)|_{\mbox{\tiny{matt}}} &=& -
   \frac{8TN_c}{\pi^2} 
   \sum_{l=1}^\infty\frac{(-1)^l}{l}\cosh \left(\frac{\mu_B l}{T}\right)
   \nonumber\\
   &\times&\sum_{i=1}^4\frac{r_im_i^2}{|b_i|^3}
   K_1\left(\frac{l|m_i|}{T}\right),
\label{condexpl}
\end{eqnarray}
where $K_1(x)$ is a Bessel function, and for convenience we have defined
$b_i=1$ for $\ i=1,2,3$, and $\ b_4=b$. 
Figure~\ref{fig2} shows the lattice QCD data for the light quark condensate as a
function of $T$~\cite{latticelatest} together with the curve obtained from the
absolute value of the sum of Eqs.~(\ref{condensatevacuum})
and~(\ref{condexpl}) for $\mu_B=0$. This parametrization gives a good
description of the condensate for the range of temperatures where the phase
transition occurs.
\section{QCD phase diagram}\label{IV}

With the parametrization of lattice data at finite $T$ and $\mu_B=0$, we proceed to extend the analysis to finite $\mu_B$. To explore the QCD phase
diagram we make use of the expressions for the light-quark condensate and of the PQCD threshold
$s_0$ that describe the chiral and deconfinement phase transitions,
respectively. Next, we compute the 
corresponding susceptibilities which  are proportional to the heat
capacities, $-\partial \langle\bar{\psi}\psi\rangle/\partial T$ and $-\partial  s_0/\partial T$.  
For a given $\mu_B$, the transition temperature is identified as the value
 $T_c$ where the heat capacity reaches a maximum. Figure~\ref{fig5} shows the
transition temperatures for the condensate and for $s_0$. These
temperatures are basically identical within a small
window of roughly $3 \;{\mbox{MeV}}$ around $T=185 \;{\mbox{MeV}}$, for all values of $\mu_B$ up to
the maximum value of $\mu_B=300 \; {\mbox{MeV}}$.

\section{Discussion and conclusions}\label{V}

In this paper we have studied the QCD phase diagram at $T\neq 0$ and $\mu_B \neq 0$ based on the behavior of the
light-quark condensate and of the PQCD threshold  as probes of  chiral symmetry restoration and deconfinement, respectively. We have shown that these quantities are related
through a QCD FESR and found that they lead to essentially equal transition
temperatures. The quark condensate, and thus the PQCD threshold, is computed
using the quark propagator in the SDE framework. We have found it convenient to use a meromorphic
parametrization of this propagator in terms of real poles and
residues. These are fixed by demanding consistency with the rainbow-ladder
truncation of SDE at $T=0$, and a good description of  
lattice QCD data for  the quark condensate at finite $T$. With this simple
scenario we have been able to extend the analysis up to baryon chemical potential  $\mu_B\simeq 300\;{\mbox{MeV}}$. From our results we can  estimate
the position of the critical end point to be $\mu_{Bc} \gtrsim 300 \;{\mbox{MeV}}$
 and $T_c\lesssim 185 \;{\mbox{MeV}}$, respectively. A more precise location of the
critical end point would require a more refined treatment of
the parametrization of the quark propagator.\\ 

\section{Acknowledgments}

We thank E. Cuautle and J. C. Arteaga for useful hints to carry out the
parametrization procedure, W. Bietenholz for pointing out useful lattice
results, P.C. Tandy for his guidance on how to parametrize the quark
propagator and extract quantities such as the condensate from this
parametrization and E. Swanson for useful suggestions. Support for this work
has been received in part by CONACyT (Mexico) under grant numbers 82230, 128534,
CIC-UMSNH 4.10 and 4.22 and PAPIIT-UNAM IN103811-3, FONDECYT 1095217 (Chile), Proyecto Anillos ACT119 (Chile), and NRF (South Africa).

\end{document}